# GENIE: Gram-Eigenmode INR Editing with Closed-Form Geometry Updates

SAMUNDRA KARKI, Iowa State University, USA
ADARSH KRISHNAMURTHY*, Iowa State University, USA
BASKAR GANAPATHYSUBRAMANIAN†, Iowa State University, USA

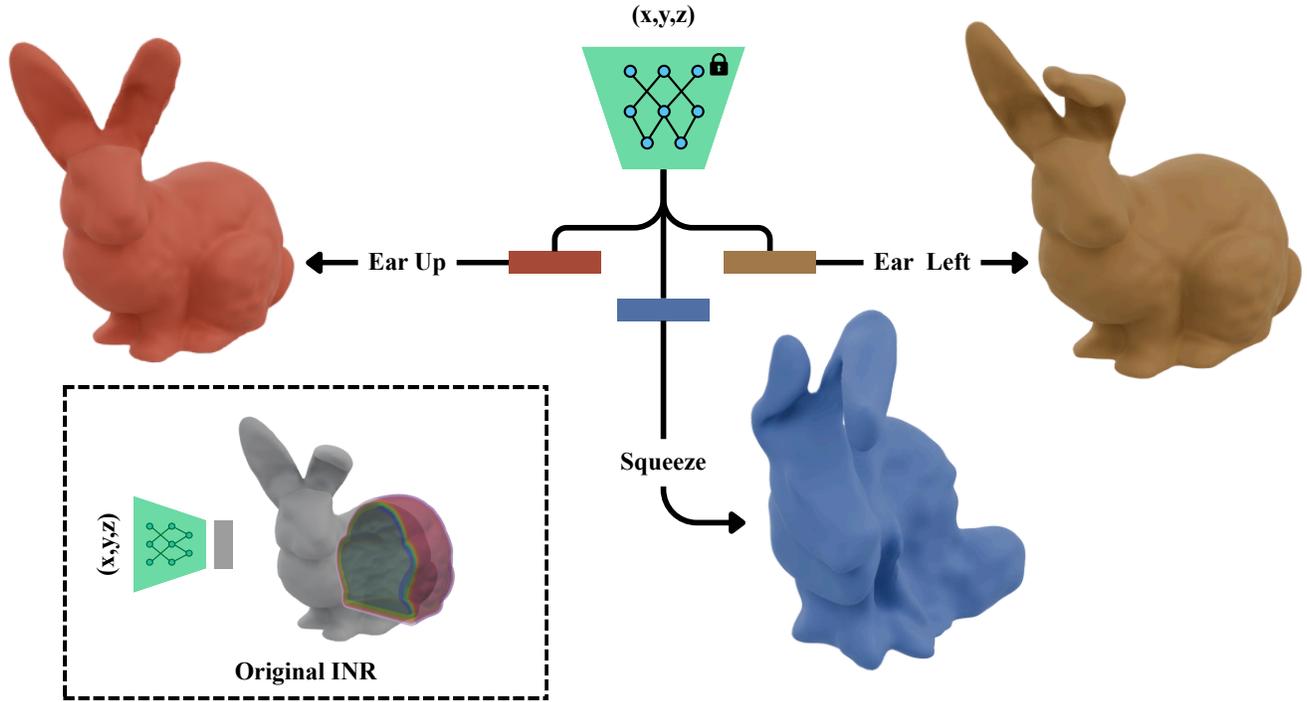

Fig. 1. **INR Editing with Gram-Eigenmode.** *GENIE* edits the last layer of a trained INR to obtain INRs with desirable edits; for example: pulling bunny ear up, pulling bunny ear to the left, or squeezing the bunny

Implicit Neural Representations (INRs) provide compact models of geometry, but it is unclear when their learned shapes can be edited without retraining. We show that the Gram operator induced by the INR's penultimate features admits deformation eigenmodes that parameterize a family of realizable edits of the SDF zero level set. A key finding is that these modes are not intrinsic to the geometry alone: they are reliably recoverable only when the Gram operator is estimated from sufficiently rich sampling distributions. We derive a single closed-form update that performs geometric edits to the INR without optimization by leveraging the deformation modes. We characterize theoretically the precise set of deformations that are feasible under this one-shot update, and show that editing is well-posed exactly within the span of these deformation modes.

*Corresponding author.
†Corresponding author.

Authors' Contact Information: Samundra Karki, Iowa State University, Ames, IA, USA, samundra@iastate.edu; Adarsh Krishnamurthy, Iowa State University, Ames, IA, USA, adarsh@iastate.edu; Baskar Ganapathysubramanian, Iowa State University, Ames, IA, USA, baskarg@iastate.edu.

## 1 Introduction

Implicit Neural Representations (INRs) have emerged as a powerful and flexible framework for encoding continuous signals such as geometry, images, and physical fields. INRs provide resolution-independent representations with strong expressivity by mapping spatial coordinates to signal values through a coordinate-based neural network. These properties have driven their rapid adoption in 3D vision, graphics, and scientific computing [3, 5, 9, 10, 15, 18]. Despite their widespread use, editing the content of a trained INR remains poorly understood. The dominant workflow relies on fine-tuning or retraining the network to achieve a desired modification, an expensive process that offers little insight into which deformations the representation can naturally support. Alternative strategies, including ad hoc manipulation of latent codes, are often heuristic, model-specific, and lack principled guarantees of *editability*. This raises a fundamental question:

Table 1. **INR architecture compatibility.** Most INR architectures have a linear final layer, making them compatible with our closed-form update.

| Model / Family | Linear? |
| --- | --- |
| SIREN [20] | ✓ |
| NeRF [15] | ✓ |
| DeepSDF [18] | ✓ |

*Given a trained INR, what geometric edits can be performed without retraining, and how can they be computed reliably?*

In this paper, we show that the answer is encoded in the last-layer Gram operator induced by the INR's penultimate features. Since the final layer of common INR architectures is linear, these features define a feature covariance operator whose eigenmodes correspond to deformation directions supported by the learned representation. Crucially, these modes are not intrinsic to the geometry alone: they depend on the sampling distribution used to estimate the Gram operator. We show that thick-band sampling and dense volumetric sampling yield a reproducible Gram operator whose spectrum consistently captures stable deformation modes. We derive a closed-form, one-shot update to the last layer that realizes geometric edits without iterative optimization by leveraging these deformation modes. We theoretically characterize the precise subspace of deformations expressible by this update, and show that INR editing is well-posed if and only if the target deformation lies within this subspace. Together, these results reveal a previously unrecognized structure in INR feature spaces and establish principled, optimization-free conditions for reliable INR editing.

*Notation.* To facilitate discussion, we introduce the following notation used throughout the paper. Let $f_\theta : \mathbb{R}^d \to \mathbb{R}$ denote an implicit neural representation (INR) of the form $f_\theta(x) = w^\top h_\phi(x) + b$, where $h_\phi(x) \in \mathbb{R}^D$ denotes the penultimate feature map, $w \in \mathbb{R}^D$ the final-layer weight vector, and $b \in \mathbb{R}$ a bias term. Given a set of $N$ sampled input points $\{x_i\}_{i=1}^N$, we stack the corresponding features row-wise to form the matrix $H \in \mathbb{R}^{N \times D}$. The associated Gram matrix used to estimate the induced operator is then $G = H^\top H \in \mathbb{R}^{D \times D}$.

## 2 Related Work

Our work lies at the intersection of three closely related research directions: (i) implicit neural representations (INRs) for geometry, (ii) editing and manipulation of learned implicit fields, and (iii) kernel and operator views of neural networks that expose analyzable structure in feature space. In this section, we review predominant paradigms for editing INRs and motivate the operator-theoretic perspective underlying our method.

*Predominant INR Editing Paradigms.* Coordinate-based neural networks (including SIREN-style sinusoidal MLPs, DeepSDF, and related implicit field models) represent continuous signals by mapping spatial coordinates to values such as signed distance, occupancy, density, or radiance. Their appeal lies in resolution independence, compactness, and their ability to couple learned geometry with rendering and simulation pipelines [3, 5, 15, 18].

Early work on INR-based geometry editing can be traced to DeepSDF, which popularized neural implicit shape representations and introduced latent-code manipulation as a mechanism for modifying geometry. This line of research gave rise to a broader literature on latent-space editing, where semantic changes are induced by perturbing or interpolating latent variables learned from shape distributions [4–6, 12, 14, 18]. While effective for coarse semantic variation, such approaches rely on access to a trained latent space and typically require retraining or iterative re-optimization to achieve precise geometric edits. Subsequent efforts explored explicit editing of INRs via optimization-based procedures, often introducing proxy representations or auxiliary networks to guide deformation toward user-specified targets [7, 22].

Some related approaches draw inspiration from level-set methods, training INRs to evolve under prescribed geometric flows or deformation rules [13, 16, 17, 19]. Boundary-sensitivity-based methods further investigate how localized perturbations propagate through implicit fields, but do not provide a clear characterization of which edits are reachable under an inexpensive parameter update [2]. A key observation emerging from several recent works is that modifying a trained INR without retraining is fundamentally constrained. Naïve parameter updates tend to induce global, unintended deformations, revealing that only a restricted class of edits can be achieved without re-optimization [21]. Existing methods mitigate this behavior empirically through careful sampling, locality heuristics, or repeated retraining, but stop short of formalizing which geometric edits are intrinsically supported by a given trained INR [4, 6, 21]. Across these paradigms, a recurring limitation is the absence of a quantitative, principled notion of *editability* for trained INRs. Our work addresses this gap by showing that editability is governed by the spectrum of the Gram operator induced by the penultimate feature representation. This operator-theoretic perspective yields a closed-form, one-shot update that succeeds *if and only if* the target deformation lies within the span of the INR's intrinsic deformation eigenmodes.

*Kernel and Operator Views of INRs.* A complementary line of work studies neural networks through kernel and operator lens. Linearization arguments (e.g., in the Neural Tangent Kernel literature) show that, around a fixed feature map, training reduces to kernel regression in a feature-induced Reproducing Kernel Hilbert Space (RKHS)[1, 8, 11]. For INRs, this viewpoint is especially natural: when the penultimate representation $h_\phi$ is held fixed, the output is linear in the last layer, and both training and editing admit closed-form expressions in terms of feature covariances.

## 3 Mathematical Background and Formulation

### 3.1 Gram Operator and Deformation Modes

Let $\mu$ denote a sampling distribution over a region of interest, such as a bounding box, a thin band around the zero level set of an SDF, or a thicker volumetric neighborhood. We draw sample points $X = \{x_i\}_{i=1}^N \sim \mu$ and evaluate the penultimate feature map $h_\phi(x) \in \mathbb{R}^D$ to construct the feature matrix

$$H = \begin{bmatrix} h_\phi(x_1)^\top \\ \vdots \\ h_\phi(x_N)^\top \end{bmatrix} \in \mathbb{R}^{N \times D}. \tag{1}$$

With the feature extractor $h_\phi$ held fixed, perturbations of the last-layer parameters $\Delta\theta \in \mathbb{R}^D$ induce output changes on the sampled points given by

$$\Delta f = H \Delta\theta, \qquad \Delta f \in \mathbb{R}^N. \tag{2}$$

Consequently, the set of realizable edits at the sampled points is precisely the column space of $H$, i.e.,

$$\mathcal{S} := \text{Range}(H) \subseteq \mathbb{R}^N.$$

While $\mathcal{S}$ characterizes the space of achievable output deformations, the geometry of this space is governed by the Gram operator

$$G := H^\top H \in \mathbb{R}^{D \times D}, \tag{3}$$

which acts in parameter (feature) space and aggregates second-order correlations between feature coordinates across the sampled domain. As the number of samples $N \to \infty$, the empirical Gram operator converges, up to normalization by $N$, to the population covariance operator

$$G_\mu = \mathbb{E}_{x \sim \mu}\left[h_\phi(x)\, h_\phi(x)^\top\right]. \tag{4}$$

This formulation makes the sampling distribution $\mu$ an explicit component of the induced operator. As a result, both the conditioning and eigenspaces of $G$ depend strongly on where $\mu$ concentrates sampling probability over the domain, a fact that plays a central role in characterizing editability in subsequent sections. The Gram operator admits an eigendecomposition, allowing us to define *Gram eigenmodes* as the eigenvectors of $G$,

$$G v_k = \lambda_k v_k, \qquad \lambda_1 \geq \lambda_2 \geq \cdots \geq 0. \tag{5}$$

Each eigenvector $v_k$ (written as *deformation modes* in subsequent texts) represents a coherent direction in feature space along which the INR can be perturbed by changing the final linear layer. Perturbing the last-layer parameters along $v_k$ induces an output variation: $\Delta f_k := H v_k \in \mathbb{R}^N$ which corresponds to a deformation evaluated at the sampled points. The squared energy of this deformation satisfies: $\|\Delta f_k\|_2^2 = v_k^\top G v_k = \lambda_k$.

Thus, the eigenvalues $\lambda_k$ quantify the strength of each deformation mode under the sampling distribution, with larger eigenvalues corresponding to more energetic and more easily realizable deformations. Importantly, these deformation modes are not guaranteed to be *geometrically meaningful* a priori; rather, they are emergent properties of the learned feature representation induced by the training procedure.

### 3.2 Closed-Form Update and One-Shot Editing

Given a target modification $y \in \mathbb{R}^N$ specified at the sampled points, we restrict editing to the final-layer parameters by solving the least-squares problem

$$\Delta\theta^\star = \arg\min_{\Delta\theta} \; \|H \Delta\theta - y\|_2^2. \tag{6}$$

The corresponding closed-form solution is

$$\Delta\theta^\star = G^{-1} H^\top y, \tag{7}$$

assuming $G$ is invertible (or using the Moore–Penrose pseudoinverse otherwise). This yields a single-shot parameter update requiring no iterative optimization. To interpret this solution, note that any target edit $y$ admits a unique orthogonal decomposition: $y = \hat{y} + r$, such that $\hat{y} \in \mathcal{S}$, $r \perp \mathcal{S}$, where $\mathcal{S} = \text{Range}(H)$ is the space of realizable edits. Since $H^\top r = 0$, the term $H^\top y$ implicitly projects the target edit onto $\mathcal{S}$. The operator $G^{-1}$ then maps this projected edit to the minimum-norm parameter update $\Delta\theta^\star$ that realizes $\hat{y}$. Equivalently, the eigenbasis of the Gram operator governs which components of the target deformation can be expressed through a last-layer update, making the spectrum of $G$ the fundamental object controlling one-shot editability.

*Editability ratio.* We quantify the extent to which a desired edit is realizable by a one-shot last-layer update via the *editability ratio*

$$\eta = \frac{\|\hat{y}\|_2^2}{\|y\|_2^2} = \frac{\|P_\mathcal{S}^{(\gamma)} y\|_2^2}{\|y\|_2^2} \in [0, 1], \tag{8}$$

where $\hat{y} = P_\mathcal{S}^{(\gamma)} y$ denotes the projection of the target edit onto the realizable subspace $\mathcal{S} = \text{Range}(H)$. An editability ratio of $\eta = 1$ indicates that the target deformation lies entirely within $\mathcal{S}$ and is therefore perfectly realizable by a single-shot update of the final layer. Conversely, $\eta = 0$ indicates that the target is orthogonal to $\mathcal{S}$ and cannot be achieved without modifying the feature extractor itself.

### 3.3 Span Enrichment

While the realizable edit subspace $\mathcal{S} = \text{Range}(H)$ captures the deformations naturally supported by a trained INR, application-driven edits may lie partially or entirely outside this span. In such cases, one-shot last-layer updates necessarily fail, as already discussed. In Section 4.4, we demonstrate that *span enrichment* through multi-head training provides a principled mechanism for enlarging $\mathcal{S}$ in controlled ways. We can encourage the INR to allocate representational capacity to additional deformation families by supervising a shared feature extractor with multiple linear heads, thereby expanding the range of edits that admit high editability ratios. As a result, deformations that are unattainable for a standard single-head INR become realizable via the same closed-form, one-shot update at test time. We illustrate this span enrichment concept in the results.

## 4 Results

Figure 2 provides an overview of our approach. We organize the paper in an inductive manner. We first study how the estimated Gram operator depends on the sampling strategy, establishing its stability and convergence properties (Section 4.1). Building on this analysis, we construct target edits within the realizable span and show that the closed-form update (Equation (7)) reliably recovers them in a single shot. We then demonstrate how multi-head training enlarges the editable span by injecting additional deformation modes during training, using a set of fundamental shapes as controlled test cases (Section 4.3, Section 4.4). Finally, we apply our framework to externally specified edits and present a comparative evaluation against relevant baselines (Section 4.5, Section 4.6). All experiments are performed with a network of 8 hidden layers, each with a dimension of 128.

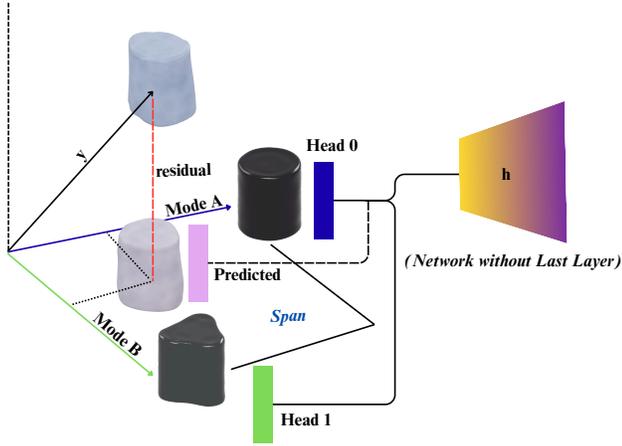

Fig. 2. **Methodology.** Diagrammatic representation of the overall framework where the axes represent two modes spanned by the hidden space of the network. The network is multi-headed, with Head 0 trained to represent the original geometry and Head 1 trained to represent some application-centric deformation. Now, we use a one-shot edit to represent a different geometry. Our methods yield a predicted head, which, combined with the hidden block, provides the best approximation within the span of the two modes.

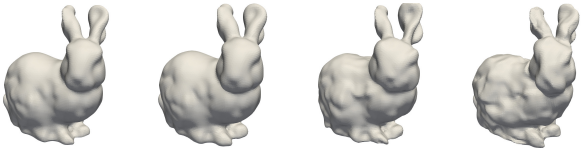

Fig. 3. **Deformation modes.** Original geometry (left) and representative deformation-span visualizations corresponding to Gram eigenmodes 0, 10, and 39.

### 4.1 Gram Matrix Stability under Sampling

We study the stability of Gram-induced deformation modes using the Stanford Bunny, represented by a standard SIREN-type MLP. Figure 3 shows representative in-span deformations obtained from the spectral decomposition of the Gram matrix $G$, illustrating deformation mode encoded by the learned feature map. We first examine the effect of the sampling region by progressively increasing the sampling band around the surface. As shown in Figure 4a, thin-band sampling yields weakly correlated deformation modes, indicating that the estimated Gram operator is sensitive to sampling noise and incomplete coverage of the feature geometry. As bandwidth increases, correlations among the leading modes increase monotonically, demonstrating that thicker sampling regions yield a more faithful and stable estimate of the Gram geometry. We next analyze convergence with respect to the number of samples under volumetric sampling. Specifically, we compute Gram operators $G^{(N)}$ for varying sample counts and compare their leading eigenspaces with a high-fidelity reference constructed from the largest available sample count. Stability is quantified via top-$k$ Gram eigenspace similarity, which is invariant to eigenvector sign ambiguities and robust to permutations within the eigenspace. As shown in Figure 4b, the leading Gram eigenspaces converge rapidly as $N$ increases, yielding reproducible and globally supported deformation modes. Taken together, these results demonstrate that consistent global deformation modes emerge only when the Gram operator is estimated from sufficiently rich volumetric samples. Both the spatial extent of the sampling region and the total number of samples play a critical role in stabilizing the Gram geometry: as sampling transitions from a thin surface band to a thicker volumetric region, the Gram estimate stabilizes, and its leading eigenspaces become reproducible, yielding consistent global deformation modes supported by the INR.

### 4.2 In-Span Deformation

We evaluate the accuracy of one-shot INR editing for *in-span* target deformations, i.e., shapes constructed as linear combinations of learned Gram eigenmodes. For each experiment, a target shape is generated by combining a small subset of eigenmodes of a trained INR, and the network is edited by updating only the final linear layer via a single closed-form solve. Figure 5 illustrates representative in-span edits for the Stanford Bunny and the Double Torus. Across randomly selected mode combinations, the edited INRs accurately reconstruct the target shapes, achieving low reconstruction error, small pointwise deviations, and low symmetric Chamfer and Hausdorff distances. Table 3 reports the metrics in the In-span edit section. These results confirm that deformations within the span of the learned Gram eigenmodes can be reliably realized via one-shot editing, yielding geometrically consistent reconstructions without iterative retraining.

### 4.3 Space Enrichment via Multi-Head Representations

To enrich the space of admissible deformations, we train a multi-headed INR in which each head corresponds to an independently learned deformation subspace. Rather than forcing all deformations to lie within a single shared feature span, the multi-head architecture allows the network to allocate representational capacity to distinct geometric modes. Figure 6 visualizes representative outputs produced by individual heads of the multi-headed INR for torus, sphere, and cylinder geometries. For each case, we use 60000 uniformly and volumetric sampled training points with 6 heads trained to 20,000 epochs with a batch size of 40,960 to minimize simple mean-squared-error loss across all heads. In each case, deformation modes are chosen to be meaningful for downstream geometric and physical tasks. For the sphere, the model is trained using solutions of the Laplace–Beltrami operator, encouraging the heads to align with intrinsic spherical harmonics. For the torus and cylinder, the heads correspond to low-frequency trigonometric modes defined over the natural angular and axial coordinates of the geometry, including uniform "breathing" modes, circumferential ovalization, axial bulging, coupled axial–angular corrugations, and twisting-like patterns. Each panel shows the deformation induced by activating a single head while keeping the shared backbone fixed. The resulting shapes exhibit diverse yet structured geometric variations, indicating that different heads specialize in capturing distinct deformation patterns and collectively enrich the admissible deformation space

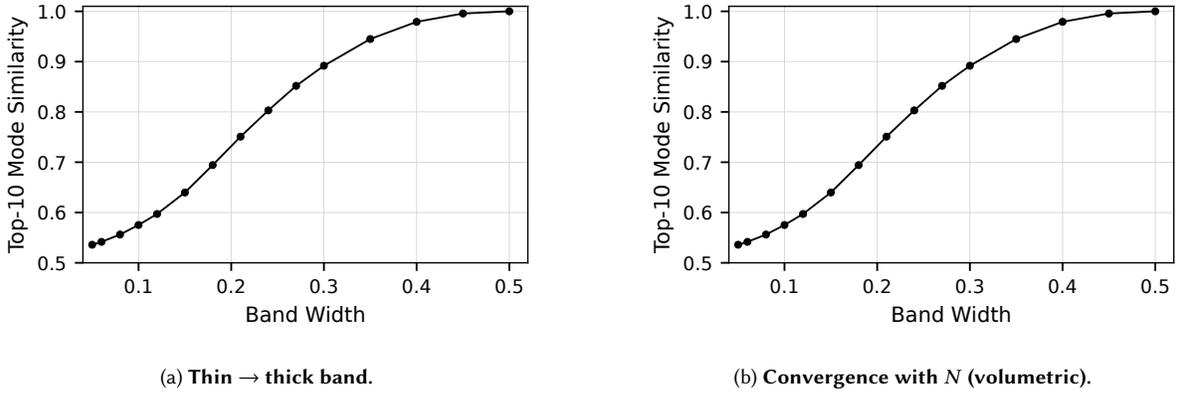

(a) **Thin → thick band.**

(b) **Convergence with $N$ (volumetric).**

Fig. 4. **Stability of Gram-induced deformation modes under sampling.** Left: increasing band width around the surface yields more consistent leading modes. Right: under volumetric sampling, the top–10 Gram eigenspace rapidly converges as the number of samples increases (largest-$N$ reference).

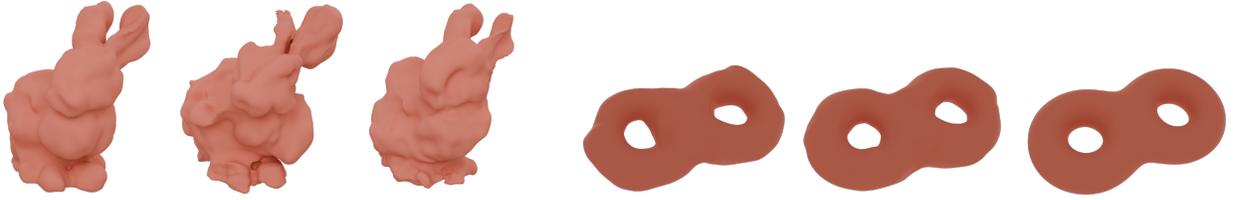

Fig. 5. **In-span one-shot INR edits.** Each target edit is synthesized as a linear combination of learned Gram eigenmodes and applied via a single closed-form update of the final linear layer.

beyond that spanned by a single shared representation. These results demonstrate that multi-headed training effectively enriches the editable space of the INR by learning multiple complementary deformation directions in parallel. In contrast to single-head models, where all edits must lie within a single Gram-induced subspace, the multi-head formulation provides a structured mechanism for expanding the set of realizable deformations that are necessary for downstream applications.

### 4.4 Quantitative evaluation of space enrichment.

To quantitatively evaluate the benefits of space enrichment, we consider a controlled editing task in which a base shape is modified using a prescribed target deformation that lies partially or entirely outside the deformation span of a single-head INR. For each base geometry, we compare edits from our method computed using (i) a single-head representation and (ii) a multi-headed representation trained to allocate deformation capacity across multiple heads. In Figure 7, we report results across five representative base shapes: ellipsoid, cylinder, sheet, torus, and sphere. For each case, the first three columns show the ground-truth target edit, followed by the reconstructed edits obtained via single-head and multi-head updates, respectively. In Table 2, we report the *editability ratio $\eta$* which measures the fraction of the target deformation captured by the admissible edit span, together with the Hausdorff distance $d_H$ between the edited shape and the ground truth averaged over a shape category. Across all shapes and deformation combinations, the multi-head formulation consistently achieves higher editability ratios ($\eta \approx 1$) and substantially lower geometric error. In contrast, single-head edits degrade when the target deformation lies outside the Gram-induced subspace learned during training, resulting in reduced editability and increased geometric distortion despite the use of an optimal least-squares update. These results demonstrate that multi-headed training effectively enriches the editable space of the INR by learning multiple complementary deformation subspaces in parallel. Crucially, this enrichment preserves the closed-form one-shot editing property, while expanding the set of deformations for which editing remains well-posed.

Table 2. **Average editing performance across base shapes.** Values are averaged over three target edits per shape. $\eta$ denotes overlap (higher is better) and $d_H$ denotes Hausdorff distance (lower is better). Multi-head editing consistently improves both overlap and boundary accuracy. (s and m in the table refer to single-head and multi-head, respectively.)

| Base Shape | $\bar{\eta}_s \uparrow$ | $\bar{d}_H^s \downarrow$ | $\bar{\eta}_m \uparrow$ | $\bar{d}_H^m \downarrow$ |
|---|---|---|---|---|
| Ellipsoid | 0.99779 | $7.53 \times 10^{-2}$ | 0.99998 | $2.89 \times 10^{-2}$ |
| Cylinder | 0.98375 | $1.27 \times 10^{-1}$ | 0.99988 | $4.64 \times 10^{-2}$ |
| Sheet | 0.98232 | $1.51 \times 10^{-1}$ | 0.99988 | $5.15 \times 10^{-2}$ |
| Sphere | 0.99543 | $6.35 \times 10^{-2}$ | 0.99996 | $2.62 \times 10^{-2}$ |
| Torus | 0.99542 | $1.52 \times 10^{-1}$ | 0.99952 | $9.51 \times 10^{-2}$ |

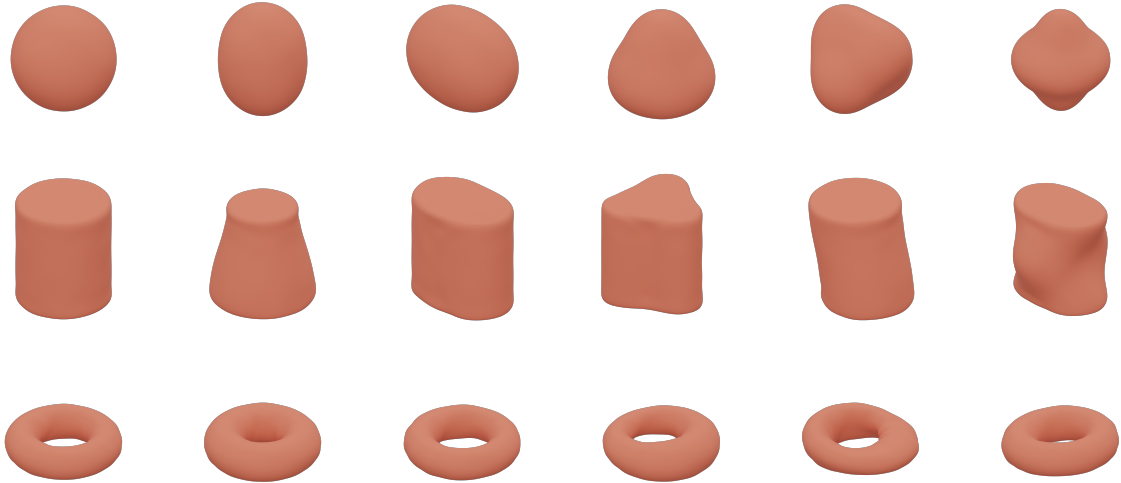

Fig. 6. **Space enrichment via multi-headed INRs.** Each row corresponds to a fixed base geometry (sphere, cylinder, torus; top to bottom), and each column shows the deformation produced by an individual head. Different heads capture distinct deformation patterns, enriching the admissible deformation space beyond a single Gram-induced subspace.

Table 3. **Quantitative evaluation of INR edits.** In-span one-shot edits achieve low Chamfer Distance (CD) and Hausdorff Distance (HD) across Bunny and Double Torus shapes. External edits—including ear lift with endpoint supervision and elastic ear pull via interpolation and extrapolation—exhibits higher error due to out-of-span deformation. Reference Head 0 provides a common baseline for both external edit types.

| Category | Edit | CD ↓ | HD ↓ |
|---|---|---|---|
| **In-span** | Bunny Edit 1 | $2.07 \times 10^{-5}$ | $3.3 \times 10^{-3}$ |
| | Bunny Edit 2 | $2.40 \times 10^{-5}$ | $1.0 \times 10^{-3}$ |
| | Bunny Edit 3 | $1.50 \times 10^{-5}$ | $5.0 \times 10^{-4}$ |
| | Dbl. Torus Edit 1 | $8.20 \times 10^{-6}$ | $2.5 \times 10^{-3}$ |
| | Dbl. Torus Edit 2 | $8.30 \times 10^{-6}$ | $7.0 \times 10^{-4}$ |
| | Dbl. Torus Edit 3 | $2.90 \times 10^{-6}$ | $1.78 \times 10^{-3}$ |
| **External** | Ear Lift (Interp. 1) | $9.21 \times 10^{-3}$ | $1.04 \times 10^{-1}$ |
| | Ear Lift (Interp. 2) | $1.06 \times 10^{-2}$ | $1.43 \times 10^{-1}$ |
| | Ear Lift (Interp. 3) | $1.21 \times 10^{-2}$ | $1.73 \times 10^{-1}$ |
| | Ear Pull (Interp. 1) | $7.77 \times 10^{-3}$ | $3.30 \times 10^{-2}$ |
| | Ear Pull (Interp. 2) | $7.66 \times 10^{-3}$ | $2.68 \times 10^{-2}$ |
| | Ear Pull (Extrap.) | $8.45 \times 10^{-3}$ | $3.43 \times 10^{-2}$ |
| | Reference Head 0 | $7.0 \times 10^{-3}$ | $2.5 \times 10^{-2}$ |

## 4.5 External Edits

We demonstrate the ability of our framework to reproduce *external geometric edits* (modifications generated entirely outside the learning pipeline) using one-shot updates on a trained multi-head INR. All edits in this section are authored manually in Blender and are neither optimized for nor constrained by the network during their construction.

*Ear lift (endpoint supervision).* Starting from a base bunny with ears in a downward pose, we construct a target edit by lifting the ears upward using Blender. To enable quantitative evaluation, we record a sequence of intermediate shapes along the editing trajectory, forming a smooth morph from the original configuration to the final lifted pose. Crucially, these intermediate shapes are *not* used during training and serve only as held-out targets. We train a multi-head INR using only the original (ears-down) geometry and the final edited (ears-up) geometry. The shared backbone learns a common geometric embedding of the bunny surface, while individual heads capture deformation modes associated with the edit endpoints. At test time, we apply a single closed-form update to the final linear layer to recover intermediate deformations. As shown in Figure 9, the recovered sequence closely matches the Blender-generated intermediates, both visually and quantitatively. As reported in the Ear lift section of the External edit in Table 3, symmetric Chamfer and Hausdorff distances remain low across the interpolation path, and views of the ear region confirm that fine-scale geometric details are preserved. These results demonstrate that, despite observing only endpoint geometries during training, the learned edit space encodes a continuous, geometrically meaningful deformation path.

*Elastic ear pull (intermediate supervision and extrapolation).* In a second experiment, we apply Blender's elastic deformation tool to pull one ear laterally, inducing a localized, highly nonlinear deformation with significant stretching and curvature. In contrast to the previous setup, we now include several intermediate meshes along the deformation trajectory during training, together with the original geometry. The final, fully deformed configuration is withheld entirely from training. After training, we again perform a one-shot update of the last-layer weights to recover unseen shapes along the deformation path. As illustrated in Figure 8, our method accurately reconstructs both withheld intermediate edits and the final extrapolated geometry. Chamfered and Hausdorff distance metrics are tabulated in the Ear pull section of the External edit in Table 3. Notably, the network recovers the large elastic displacement and localized surface distortion of the ear, even though the extrapolated shape was never observed during training. This highlights a complementary regime in which intermediate supervision enables reliable extrapolation along deformation trajectories within the learned edit subspace.

### 4.6 Comparative Study

Table 4 reports the average editing performance across six bunny shape interpolations, comparing our proposed one-shot update against displacement-based and iterative baselines. Our method, which performs a single closed-form ridge update on the last layer using signed distance residuals, achieves the lowest Chamfer and Hausdorff errors while being orders of magnitude faster than all competing approaches. We observe that the Boundary Sensitivity [2] approach, implemented here as a last-layer ridge update based on its linearized normal-displacement formulation, incurs substantially higher computational cost and yields worse geometric accuracy. Iterative gradient descent on the last layer improves over Boundary Sensitivity in terms of runtime and accuracy but remains notably slower and less accurate than the proposed one-shot update. Finally, the unconstrained optimization of all network parameters (**GD-BS (all)**) results in catastrophic geometric degradation. We observe that allowing deep updates against the linearized boundary-sensitivity objective causes the network to overfit to the linearization approximation, leading to high-frequency surface noise and a breakdown of the underlying signed distance field structure. This confirms that unrestricted fine-tuning induces *manifold fracturing*, whereas restricting updates to the Gram-induced subspace preserves global geometric continuity. Overall, these results highlight that reliable edits are best achieved through structured, spectral updates aligned with the representation, rather than through deep iterative optimization.

### 5 Conclusions

We presented *GENIE*, a method for editing implicit neural representations (INRs) through the spectral geometry of their learned feature space. By analyzing the Gram operator induced by the penultimate feature extractor, we identified a principled subspace of admissible deformations that can be realized efficiently via one-shot last-layer updates. We showed that this subspace captures meaningful geometric variation and that multi-head training systematically enriches the editable space, enabling a broader class of application-driven edits without iterative retraining. Taken together, these results suggest that even relatively simple MLP-based INRs can serve as effective, controllable geometry layers in scientific computing workflows, where rapid, low-cost geometric updates are often required (e.g., design exploration, sensitivity studies, and parametric PDE simulations). The ability to enact fast, structured edits without full retraining makes such representations particularly attractive for interactive and simulation-in-the-loop settings.

Table 4. **Average editing performance.** Editing performance over six bunny shape edits, averaged across six target interpolations. **GENIE** performs a single closed-form last-layer update from SDF residuals. **BS-L (last)** applies the boundary sensitivity method of Berzins et al. [2] restricted to the last layer. **GD-SDF (last)** uses gradient descent to minimize SDF regression loss on the last layer. **GD-BS (last)** performs gradient descent on a boundary-sensitivity-inspired proxy objective for the last layer, while **GD-BS (all)** updates all network parameters. All methods use the same number of sampling points and 400 gradient descent steps where applicable. Lower is better for all metrics.

| Method | Time (s) ↓ | CD ↓ | HD ↓ |
|---|---|---|---|
| **GENIE (Ours)** | **0.0350** | **0.0079** | **0.0339** |
| BS-L (last) | 17.400 | 0.0277 | 0.1890 |
| GD-SDF (last) | 1.4700 | 0.0099 | 0.0657 |
| GD-BS (last) | 5.0700 | 0.0378 | 0.1940 |
| GD-BS (all) | 7.9400 | 0.6010 | 1.4090 |

At the same time, our study highlights important limitations. Precise control requires prior knowledge of the span of admissible deformations supported by a trained representation; edits outside this span remain unattainable through inexpensive last-layer updates. Moreover, while multi-head training expands the editable space, we have not yet quantified how shared training affects the fidelity of the baseline INR, nor how competing deformation modes interact. Characterizing the trade-off between editability and representation accuracy is an important direction for future work.

Beyond these questions, our results open two concrete avenues. First, we aim to construct *a priori* deformation spans that encode design intent and constraints, and then search within these spans for optimal geometries while retaining a fixed, reusable INR backbone. This would connect GENIE-style editing to simulation-driven design pipelines in which geometry updates must remain compatible with downstream solvers (e.g., INR-based variants of immersed-boundary or embedded-boundary methods). Second, we plan to investigate the use of *a priori* defined INRs as geometry and motion parameterizations for coupled multiphysics, including fluid–structure interaction (FSI). In this setting, the editable subspace provides a low-dimensional, structured control space for geometry evolution, potentially enabling stable coupling, efficient time stepping, and gradient-based optimization with the INR serving as a differentiable interface between evolving geometry and the PDE discretization.

8 • Samundra Karki et al.

## References

[1] Sanjeev Arora, Simon S Du, Wei Hu, Zhiyuan Li, Russ R Salakhutdinov, and Ruosong Wang. 2019. On Exact Computation with an Infinitely Wide Neural Net. In *Advances in Neural Information Processing Systems*, H. Wallach, H. Larochelle, A. Beygelzimer, F. d'Alché-Buc, E. Fox, and R. Garnett (Eds.), Vol. 32. Curran Associates, Inc., Vancouver, 1–12. https://proceedings.neurips.cc/paper_files/paper/2019/file/dbc4d84bfcfe2284ba11beffb853a8c4-Paper.pdf

[2] Arturs Berzins, Moritz Ibing, and Leif Kobbelt. 2023. Neural Implicit Shape Editing using Boundary Sensitivity. In *International Conference on Learning Representations (ICLR)*. Curran Associates, Kigali, 1–12. https://openreview.net/forum?id=CMPIBjmhpo

[3] Julian Chibane, Thiemo Alldieck, and Gerard Pons-Moll. 2020. Implicit functions in feature space for 3D shape reconstruction and completion. In *Proceedings of the IEEE/CVF Conference on Computer Vision and Pattern Recognition (CVPR)*. IEEE, virtual, 6970–6981.

[4] Tim Elsner, Moritz Ibing, Victor Czech, Julius Nehring-Wirxel, and Leif Kobbelt. 2021. Intuitive Shape Editing in Latent Space. arXiv:2111.12488 [cs.GR]

[5] Amos Gropp, Lior Yariv, Niv Haim, Matan Atzmon, and Yaron Lipman. 2020. Implicit geometric regularization for learning shapes. In *Proceedings of the 37th International Conference on Machine Learning (ICML'20)*. JMLR.org, virtual, Article 355, 11 pages.

[6] Zekun Hao, Hadar Averbuch-Elor, Noah Snavely, and Serge Belongie. 2020. DualSDF: Semantic Shape Manipulation Using a Two-Level Representation. In *Proceedings of the IEEE/CVF Conference on Computer Vision and Pattern Recognition (CVPR)*. IEEE, virtual, 1–12.

[7] Amir Hertz, Or Perel, Raja Giryes, Olga Sorkine-Hornung, and Daniel Cohen-Or. 2022. SPAGHETTI: editing implicit shapes through part aware generation. *ACM Trans. Graph.* 41, 4, Article 106 (July 2022), 20 pages. doi:10.1145/3528223.3530084

[8] Arthur Jacot, Franck Gabriel, and Clement Hongler. 2018. Neural Tangent Kernel: Convergence and Generalization in Neural Networks. In *Advances in Neural Information Processing Systems*, S. Bengio, H. Wallach, H. Larochelle, K. Grauman, N. Cesa-Bianchi, and R. Garnett (Eds.), Vol. 31. Curran Associates, Inc., Montréal. https://proceedings.neurips.cc/paper_files/paper/2018/file/5a4be1fa34e62bb8a6ec6b91d2462f5a-Paper.pdf

[9] Samundra Karki, Ming-Chen Hsu, Adarsh Krishnamurthy, and Baskar Ganapathysubramanian. 2026. Mechanics simulation with Implicit Neural Representations of complex geometries. *Computer-Aided Design* 190 (2026), 103978. doi:10.1016/j.cad.2025.103978

[10] Samundra Karki, Mehdi Shadkhah, Cheng-Hau Yang, Aditya Balu, Guglielmo Scovazzi, Adarsh Krishnamurthy, and Baskar Ganapathysubramanian. 2025. Direct flow simulations with implicit neural representation of complex geometry. *Computer Methods in Applied Mechanics and Engineering* 446 (2025), 118248.

[11] Jaehoon Lee, Lechao Xiao, Samuel Schoenholz, Yasaman Bahri, Roman Novak, Jascha Sohl-Dickstein, and Jeffrey Pennington. 2019. Wide Neural Networks of Any Depth Evolve as Linear Models Under Gradient Descent. In *Advances in Neural Information Processing Systems*, H. Wallach, H. Larochelle, A. Beygelzimer, F. d'Alché-Buc, E. Fox, and R. Garnett (Eds.), Vol. 32. Curran Associates, Inc., Vancouver. https://proceedings.neurips.cc/paper_files/paper/2019/file/0d1a9651497a38d8b1c3871c84528bd4-Paper.pdf

[12] Hsueh-Ti Derek Liu, Francis Williams, Alec Jacobson, Sanja Fidler, and Or Litany. 2022. Learning Smooth Neural Functions via Lipschitz Regularization. In *ACM SIGGRAPH 2022 Conference Proceedings* (Vancouver, BC, Canada) *(SIGGRAPH '22)*. Association for Computing Machinery, New York, NY, USA, Article 31, 13 pages. doi:10.1145/3528233.3530713

[13] Ishit Mehta, Manmohan Chandraker, and Ravi Ramamoorthi. 2022. A Level Set Theory for Neural Implicit Evolution Under Explicit Flows. In *Computer Vision – ECCV 2022*, Shai Avidan, Gabriel Brostow, Moustapha Cissé, Giovanni Maria Farinella, and Tal Hassner (Eds.). Springer Nature Switzerland, Cham, 711–729.

[14] Lars Mescheder, Michael Oechsle, Michael Niemeyer, Sebastian Nowozin, and Andreas Geiger. 2019. Occupancy Networks: Learning 3D Reconstruction in Function Space. In *Proceedings of the IEEE/CVF Conference on Computer Vision and Pattern Recognition (CVPR)*. IEEE, Long Beach, 1–12.

[15] Ben Mildenhall, Pratul P. Srinivasan, Matthew Tancik, Jonathan T. Barron, Ravi Ramamoorthi, and Ren Ng. 2021. NeRF: representing scenes as neural radiance fields for view synthesis. *Commun. ACM* 65, 1 (Dec. 2021), 99–106. doi:10.1145/3503250

[16] Michael Niemeyer, Lars Mescheder, Michael Oechsle, and Andreas Geiger. 2019. Occupancy Flow: 4D Reconstruction by Learning Particle Dynamics. In *Proceedings of the IEEE/CVF International Conference on Computer Vision (ICCV)*. IEEE, Seoul, 1–12.

[17] Tiago Novello, Vinicius da Silva, Guilherme Schardong, Luiz Schirmer, Helio Lopes, and Luiz Velho. 2023. Neural Implicit Surface Evolution. In *Proceedings of the IEEE/CVF International Conference on Computer Vision (ICCV)*. IEEE, Paris, 14279–14289.

[18] Jeong Joon Park, Peter Florence, Julian Straub, Richard Newcombe, and Steven Lovegrove. 2019. DeepSDF: Learning Continuous Signed Distance Functions for Shape Representation. In *Proceedings of the IEEE/CVF Conference on Computer Vision and Pattern Recognition (CVPR)*. IEEE, Long Beach, 1–12.

[19] Lu Sang, Zehranaz Canfes, Dongliang Cao, Florian Bernard, and Daniel Cremers. 2025. Implicit Neural Surface Deformation with Explicit Velocity Fields. In *The Thirteenth International Conference on Learning Representations*. Curran Assoc, Singapore, 1–12.

[20] Vincent Sitzmann, Julien Martel, Alexander Bergman, David Lindell, and Gordon Wetzstein. 2020. Implicit Neural Representations with Periodic Activation Functions. In *Advances in Neural Information Processing Systems*, H. Larochelle, M. Ranzato, R. Hadsell, M.F. Balcan, and H. Lin (Eds.), Vol. 33. Curran Associates, Inc., virtual, 7462–7473. https://proceedings.neurips.cc/paper_files/paper/2020/file/53c04118df112c13a8c34b38343b9c10-Paper.pdf

[21] Petros Tzathas, Petros Maragos, and Anastasios Roussos. 2023. 3D Neural Sculpting (3DNS): Editing Neural Signed Distance Functions. In *Proceedings of the IEEE/CVF Winter Conference on Applications of Computer Vision (WACV)*. Curran Assoc., Waikoloa, 4521–4530.

[22] Guandao Yang, Serge Belongie, Bharath Hariharan, and Vladlen Koltun. 2021. Geometry Processing with Neural Fields. In *Advances in Neural Information Processing Systems*, M. Ranzato, A. Beygelzimer, Y. Dauphin, P.S. Liang, and J. Wortman Vaughan (Eds.), Vol. 34. Curran Associates, Inc., virtual, 22483–22497. https://proceedings.neurips.cc/paper_files/paper/2021/file/bd686fd640be98efaae0091fa301e613-Paper.pdf

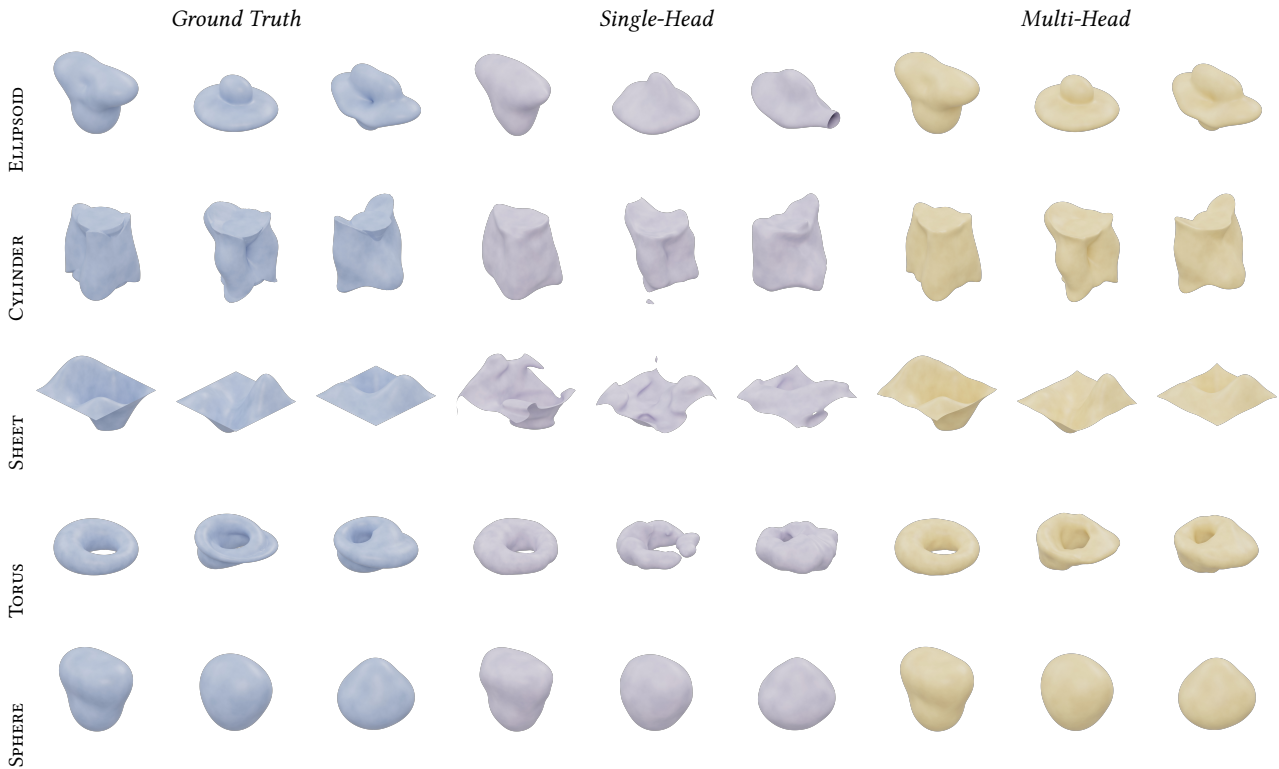

Fig. 7. **Span Enrichment Visualization.** Qualitative comparison of single-head vs multi-head edits across multiple shapes. Each row displays three Ground Truth edits from the original shape. Followed by results of Single Head Edits and Multi-Head Edits.

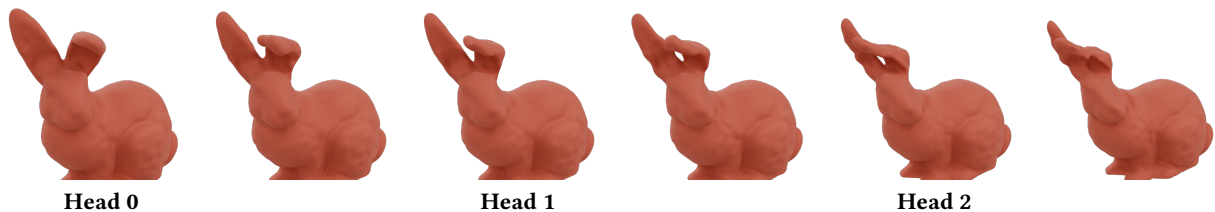

Fig. 8. **Elastic Ear-pull edit**. Bunny ear movement towards left progression across multiple heads. The figures between the trained heads are interpolations followed by an extrapolation at the end.

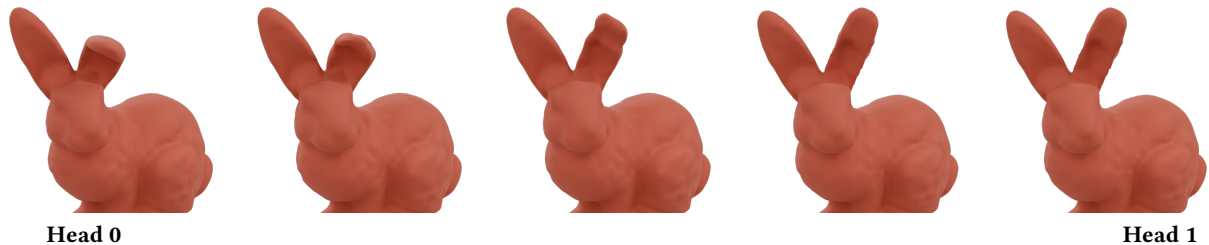

Fig. 9. **Ear-Up Edit.** One-shot interpolation between two heads for the Bunny ear-up edit. Left to right, the five shapes correspond to Head 0, three intermediate interpolants, and Head 1.